\documentclass[twocolumn]{aastex6}

\usepackage{epsfig}
\usepackage{graphicx}
\usepackage{url}
\usepackage{natbib}
\usepackage{float}
\usepackage{multirow}
\usepackage{cases}
\usepackage{color}

\newcommand{\kms}{km\,s$^{-1}$}

\newcommand{\vR}{$v_R$}

\newcommand{\vPhi}{$v_\phi$}
\newcommand {\Usun}{{U$_{\odot}$}}
\newcommand {\Wsun}{{W$_{\odot}$}}
\newcommand {\Vsun}{{V$_{\odot}$}}

\newcommand{\masyr}{mas yr$^{-1}$}

\begin{document}


\title{The local spiral arm in the LAMOST-\emph{Gaia} common stars?}


\author{Chao Liu\altaffilmark{1}}
\affil{Key Laboratory for Optical Astronomy, National Astronomical Observatories, Chinese Academy of Sciences, Beijing 100012, China}
\author{You-Gang Wang}
\affil{Key Laboratory of Computational Astrophysics, National Astronomical Observatories, Chinese Academy of Sciences, Beijing, 
100012, China}
\author{Juntai Shen, Zhao-Yu Li, Yu-Jing Qin}
\affil{Key Laboratory for Research in Galaxies and Cosmology, Shanghai Astronomical Observatory, Chinese Academy of Sciences, Shanghai, 200030, China}
\author{Yonghui Hou, Yuefei Wang, Yong Zhang}
\affil{Nanjing Institute of Astronomical Optics \& Technology, National Astronomical Observatories, Chinese Academy of Sciences, Nanjing 210042, China}
\and
\author{Zihuang Cao, Yue Wu}
\affil{Key Laboratory of Optical Astronomy, National Astronomical Observatories, Chinese Academy of Sciences, Beijing 100012, China}

\altaffiltext{1}{liuchao@nao.cas.cn}

\begin{abstract}
Using the LAMOST-\emph{Gaia} common stars, we demonstrate that the in-plane velocity field for the nearby young stars are significantly different from that for the old ones. For the young stars, the probably perturbed velocities similar to the old population are mostly removed from the velocity maps in the $X$--$Y$ plane. The residual velocity field shows that the young stars consistently move along $Y$ with faster $v_\phi$ at the trailing side of the local arm, while at the leading side, they move slower in azimuth direction. At both sides, the young stars averagely move inward with $v_R$ of $-5\sim-3$\,km s$^{-1}$. The divergence of the velocity in $Y$ direction implies that the young stars are associated with a density wave nearby the local arm. We therefore suggest that the young stars may reflect the formation of the local spiral arm by correlating themselves with a density wave. The range of the age for the young stars is around 2\,Gyr, which is sensible since the transient spiral arm can sustain that long. We also point out that alternative explanations of the peculiar velocity field for the young population cannot be ruled out only from this observed data.
\end{abstract}




\section{Introduction}\label{sect:intro}
The spiral arms are very common features in disk galaxies and also found in our Galaxy~\citep{kerr1957,benjamin2005}. However, the origin of these complicated structures is not conclusively solved, although many theories are proposed~\citep[see][]{binney2008,shu2016}. \citet{lin1964} suggested the spiral structures are long-lived density wave. However, this was challenged by observations for external galaxies~\citep{foyle2011}. \citet{toomre1972} found that dynamical tidal interaction can induce spiral structures in a galaxy. But this is hard to explain why galaxies not experienced major mergers also show spiral arms, such as the Milky Way. \citet{sellwood1984} argued that the spiral structures are short-lived and formed from the instability. \citet{sellwood2014} further explained that the nature of the recurrent spiral arms are the overlapping of multiple spiral modes. \citet{kawata2014}, on the other hand, claimed that the spiral arms are corotating, i.e. their pattern speeds equal to the circular speeds.

Because the dynamical properties of the spiral structures in the Milky Way is hard to be directly detected, most of the observational works turn to understand the spiral structure through the perturbed stellar velocity distributions~\citep{quillen2011,siebert2012,faure2014}. In recent years, many observations revealed that the nearby stars display quite complicated peculiar motions, which may be the result of perturbations, in radial and vertical directions~\citep[][etc.]{siebert2011,williams2013,carlin2013,sun2015}. However, the perturbed velocities (also called bulk motion, asymmetric motion, streaming velocity, or peculiar velocity in different studies) may be either induced by the spiral structures, or other non-axisymmetric features, e.g. the Galactic bar or the merging dwarf galaxies~\citep{bovy2015,grand2015,gomez2013}. There are also some works constraining the forming mechanisms of the spiral structures with the comparison of the locations of the gaseous and stellar arms~\citep{hou2015}.

Recently, \citet{tian2015} showed that the young stars have different asymmetric motions with the old stars in $U$ and $W$. \citet{tian2016} further compared the radial variation of \vR\ and \vPhi\ between the young and relatively old red clump stars from the LAMOST survey and found that the radial oscillations of the velocities for the two populations show similar long-wave mode but different short-wave mode. These observational facts hint that the young stars may contain different types of peculiar velocities. They may either retain some memory of the kinematical features of their birthplace, or directly reflect the kinematics of the spiral arms, or more sensitive to the perturbations than older populations due to their more circular stellar orbits. In order to investigate the role played by the spiral arms in the peculiar velocities, two-dimensional in-plane velocity field, rather than one-dimensional radial variation of the velocities, is required. Subsequently, the accurate astrometric data, combined with the line-of-sight velocity measured from the stellar spectra, are needed. 

As September 14, 2016, the \emph{Gaia} survey~\citep{gaiamission} published its first data release~\citep{gaiasummary}, including new proper motion estimates with accuracy of $\sim1$\,\masyr\ for 2 million bright stars~\citep{lindegren2016}. The \emph{Gaia} DR1, combined with the spectroscopic data from the LAMOST DR3 data, allows us to map the three dimensional velocities of the stars within $\sim1$\,kpc around the Sun with high accuracy.

In this letter, we study the in-plane peculiar motions for the young stars in the solar neighborhood using the latest astrometric data from \emph{Gaia}. The data selection and processing are described in Section~\ref{sect:data}. The result and discussions are illustrated in Section~\ref{sect:result}. And a brief conclusion is drawn in the last section.

\section{Data}\label{sect:data}

\begin{figure}[htbp]
\plotone{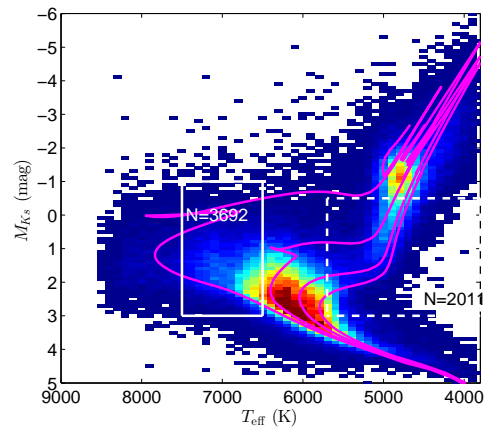}
\caption{This figure illustrates the color coded logarithmic stellar density in the $M_{Ks}$ vs. $T_{\rm eff}$ plane for the LAMOST-\emph{Gaia} stars with errors of the  distance smaller than 30\%. The left white solid rectangle indicates the selection of the F-type stars as tracers of the young population and the right white dashed rectangle marks the old K-type sub-giant branch stars as the control sample. The magenta lines indicate the PARSEC isochrone tracks with ages of 1, 3, 6, and 10\,Gyr from top to bottom, respectively, at the solar metallicity.}\label{fig:HRD}	
\end{figure}

The data used in this work are from two surveys, the stellar parameters and the line-of-sight velocities are from the LAMOST DR3 data~\citep{cui2012,zhao2012,deng2012} and the parallaxes and proper motions are from the \emph{Gaia} DR1~\citep{gaiasummary}. We adopt the distance estimated by \citet{astraatmadja2016}, who applied a Bayesian model to derive the distance from the parallax taking into account the Milky Way prior and systematic uncertainties in \emph{Gaia} catalog. There are totally more than 190\,000 common stars in the two catalogs. From the LAMOST derived stellar parameters~\citep{wu2014}, we select about 15\,000 young F-type stars with $6500<T_{\rm eff}<7500$\,K, error of distance from ~\citet{astraatmadja2016} smaller than 30\% (including the systematic error of the \emph{Gaia} parallax), and absolute magnitude in $K_s$ band between -1 and 3\,mag\footnote{The absolute magnitude in $K_s$ band is derived from the distance provided by~\citet{astraatmadja2016}, the apparent $K_s$ magnitude from 2MASS, and the derived interstellar extinction in $K_s$ band according to~\citet{majewski2011}.}. Because in this work we focus on the disk, we cut the range of vertical distance to the Galactic mid-plane with $|Z|<0.3$\,kpc. Meanwhile, we cut the distance with $D<0.6$\,kpc  to ensure the completeness of the data. Then we obtain 6\,822 F-type stars as the tracers of the young population. In order to characterize the velocities for the young population, we also select 5\,088 K-type sub-giant branch (SGB) stars with $3800<T_{\rm eff}<5700$\,K and $-0.5<M_{K_s}<3$ as the control sample. Figure~\ref{fig:HRD} shows the HR diagram of the LAMOST-\emph{Gaia} common stars overlapped with the PARSEC isochrones~\citep{bressan2012} with age of 1, 3, 6, and 10\,Gyr at solar metallicity (magenta lines). It is seen that most of the selected F-type stars (left white solid box) are around 1--3\,Gyr, while the K-type SGB control sample (the right white dashed box) are concentrated in the age around 3--6\,Gyr.

There are some stars with supersolar metallicity in both the young and old populations. According to \citet{kordopatis2015} and \citet{liu2015}, these very metal-rich stars are likely formed from the inner disk. In order to make sure that the sample are local, we remove the stars with [Fe/H]$>-0.05$\,dex. We also cut out a few stars with [Fe/H]$<-0.6$ to exclude most of the contaminations, e.g. the blue straggler stars, from the thick disk and stellar halo populations. After the cut in metallicity, the number of the rest F-type young stars is 3\,692 and the number of the rest K-type SGB stars is 2\,011.

\begin{figure}[htb]
\centering
\includegraphics[scale=0.5,trim={0.5cm 0 0 0},clip]{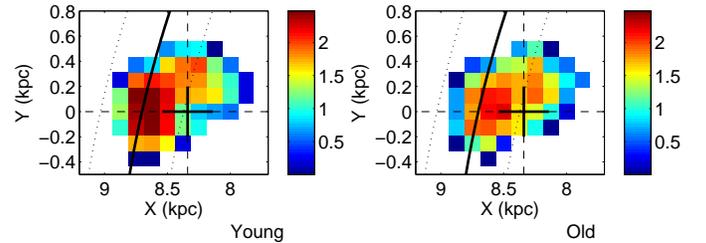}
\caption{The spatial distributions for the young and old samples in left and right panels, respectively. The color codes the logarithmic stellar density in $X$--$Y$ plane. The location of the Sun, which is marked as large black cross, is at ($X$, $Y$)=(8.34, 0)\,kpc, the Galactic center is at the origin in the right side of the plots, and the direction of $Y$ is aligned with the direction of the rotation of the disk. The black solid line and the two black dotted lines crossing from top to bottom in each panel indicate the local spiral arm and its width, respectively.}\label{fig:N}
\end{figure}

We adopt that the solar motion w.r.t. the local standard of rest is (\Usun, \Vsun, \Wsun)=(9.58, 10.52, 7.01)\,\kms\ \citep{tian2015}, the distance from the Sun to the Galactic center is $R_0=8.34$\,kpc~\citep{reid2014}, and the local circular speed is 238\,\kms~\citep{schoenrich2012}. Figure~\ref{fig:N} shows the spatial distribution of the young (left panel) and old (right panel) stars in $X$--$Y$ plane, in which $X$ points from the Galactic center to the outskirt of the disk through the Sun and $Y$ points to the direction of the rotation of the disk. Noted that this right-hand coordinates does not fit the usual direction to look at the Milky Way, i.e. looking it from north Galactic pole. Hence, in all rest figures in this letter, we flip the direction of $X$-axis so that the Milky Way is looked from the north Galactic pole and the direction of the rotation is upward. The samples are located within 0.6\,kpc around the Sun. Thus, $Y$ is approximately equivalent with the azimuth angle.

\section{Result \& Discussions}\label{sect:result}
\begin{figure*}[htb]
\centering
\begin{minipage}{18cm}
\centering
\includegraphics[scale=0.55,trim={1.5cm 3.5cm 0 5cm},clip]{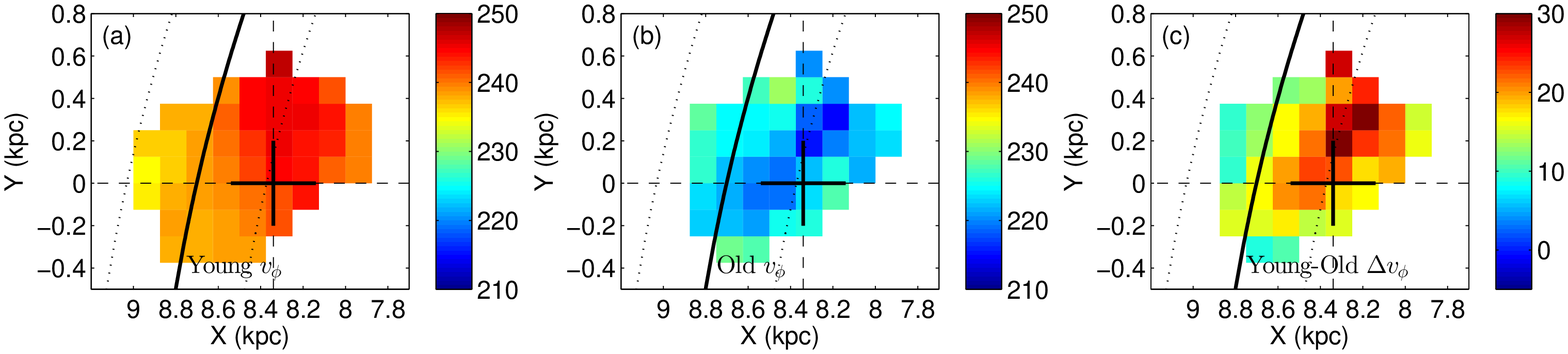}\vspace{0pt}
\includegraphics[scale=0.55,trim={1.5cm 3.5cm 0 5cm},clip]{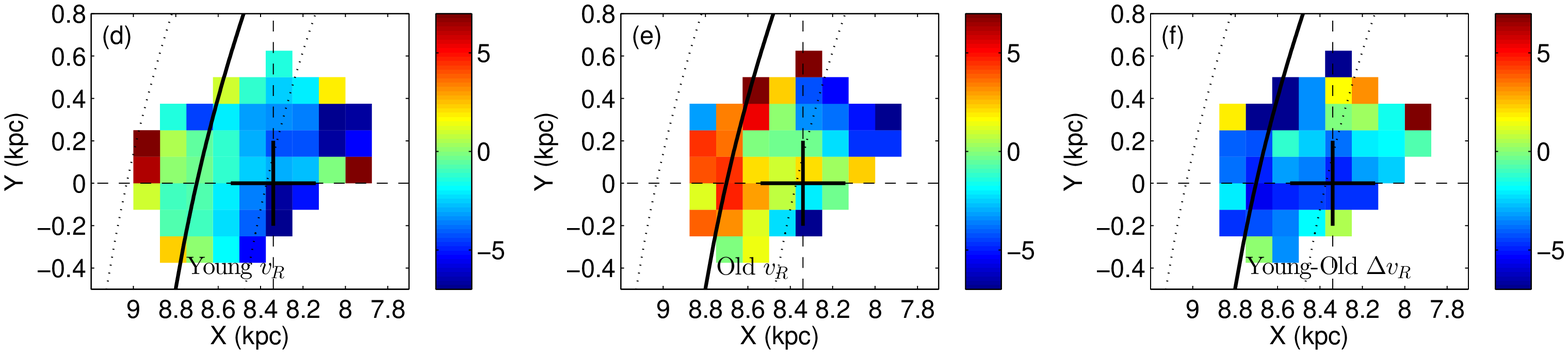}\vspace{0pt}
\end{minipage}
\caption{The top row shows the maps of the median \vPhi\ in the $X$--$Y$ plane for the F-type stars (young) and K-type SGB stars (old) in panels (a) and (b), respectively. The location of the Sun is at the black crosses. The black solid line and the two black dotted lines crossing from top to bottom in each panel indicate the local spiral arm and its width, respectively. Panel (c) shows the residual map of the subtraction of panel (b) from panel (a). The bottom row shows the maps of the median \vR\ in the $X$--$Y$ plane for the F-type stars (young) and K-type SGB stars (old) in panels (d) and (e), respectively. Panel (f) shows the residual map of the panel (d) subtracted by panel (e).}\label{fig:VPhi}
\end{figure*}

Figure~\ref{fig:VPhi} shows the maps of the median azimuthal (\vPhi) and radial (\vR) velocities in $X$--$Y$ plane for the young and old stars, respectively. The median velocity values are provided only for the bins containing more than 20 stars to ensure the good statistics. The bin size is 0.125$\times$0.125\,kpc. No smoothing technique is applied to these maps. The local spiral arm derived from the astrometry of the masers~\citep{reid2014,xu2016} is superposed to the figure (the center of the arm is represented by the solid lines and the width of it by dotted lines). 

From panel (a) it is seen that the map of the median \vPhi\ for the young population displays clear declining trend with increasing $X$. As a comparison, panel (b) shows the map of \vPhi\ for the old K-type SGB stars. It shows a ridge located from (8.9,-0.2) to (8.2,+0.4)\,kpc with lower \vPhi\ of about 220\,\kms. Outside the ridge, \vPhi\ increases to around 225\,\kms. This is substantially different with the velocity gradient for the young stars shown in panel (a). The pattern of \vPhi\ for the old stars seems correlated with the direction of the local arm. This hints that the velocity pattern may reflect the perturbation induced by the spiral arm.

We then look at the map of \vR\ for the young population in Figure~\ref{fig:VPhi} (d). It shows that the young stars move inward with \vR$\sim-3$\,\kms\ at $R<8.5$\,kpc, while their \vR\ is at around zero at $R>8.5$\,kpc. Looking at the map of \vR\ for the old population in panel (e), we find, roughly, \vR\ is around $+5$\,\kms\ at $R>8.5$\,kpc and about zero or even smaller than zero at $R<8.5$\,kpc. The trend of the gradients of \vR\ for the young and old populations are similar. However, the values of \vR\ are different such that the old populations show larger $v_R$ than the young one by around 5\,\kms. It is noted that the error of \vR\ for our sample is about 5\,\kms, which leads to the order of $\sim1$\,\kms\ in each $X$--$Y$ bin since each bin contains at least 20 stars. Because the uncertainty of \vR\ in each bin is comparable with the local variations in the \vR\ map, these variations between neighboring bins in panels (d) and (e) are likely dominated by the statistical fluctuation. However, The global radial trend displayed in both the young and old populations may not be due to the velocity uncertainties, but is likely a true feature since lots of bins demonstrate this trend together.

The old stars should be completely relaxed. Hence, the peculiar velocity shown in the old stars is most likely due to the perturbations induced by the spiral arm or the bar. In principle, if the kinematics for the old stars suffers from the perturbations, the young population should be affected in similar way with similar or even more intensive amplitudes because that they are more kinematically cold and thus easier to be perturbed. Therefore, the significant difference in the map of velocities between the young and old populations in panels (a), (d), (b), and (e) of Figure~\ref{fig:VPhi} implies that the young stars may be driven by two different mechanisms: 1) the perturbations same as those also affecting the old stars and 2) a special mechanism plays nothing with the old populations, but only influences the young stars. Then we can remove most of the perturbed velocities due to the first driver by subtracting the maps of velocities for the old stars from those for the young stars in the $X$--$Y$ plane.

The residual velocity maps are shown in Figure~\ref{fig:VPhi} (c) and (f). In most parts of the $X$--$Y$ plane, the residual \vPhi\ in panel (c) is positive from about 10 to 30\,\kms. The positive residual values can be naturally explained by the different asymmetric drift, which is smaller for the young populations than for the old ones. The residual \vPhi\ also displays a clear gradient from bottom-right to top-left. The direction of the gradient is approximately parallel with the norm direction of the local arm. It shows that the young stars move along the azimuthal direction faster in the trailing side of the local arm, while they moves slower by 20\,\kms\ in the leading side. 

The residual \vR\ in panel (f) displays negative values around $-5\sim-3$\,\kms\ with some local variations. Consider the uncertainty of \vR\ in each bin is around 1\,\kms, the local features with variation of $1\sim2$\,\kms\ may be owe to the arbitrary fluctuation. However, even taking into account the uncertainties, panel (f) still shows that, averagely, the \vR\ of the young population is about $3\sim5$\,\kms\ smaller than that of the old one.

\begin{figure}[htbp]
\centering
\begin{minipage}{9cm}
\centering
\includegraphics[scale=0.5,trim={0cm 0 0 0},clip]{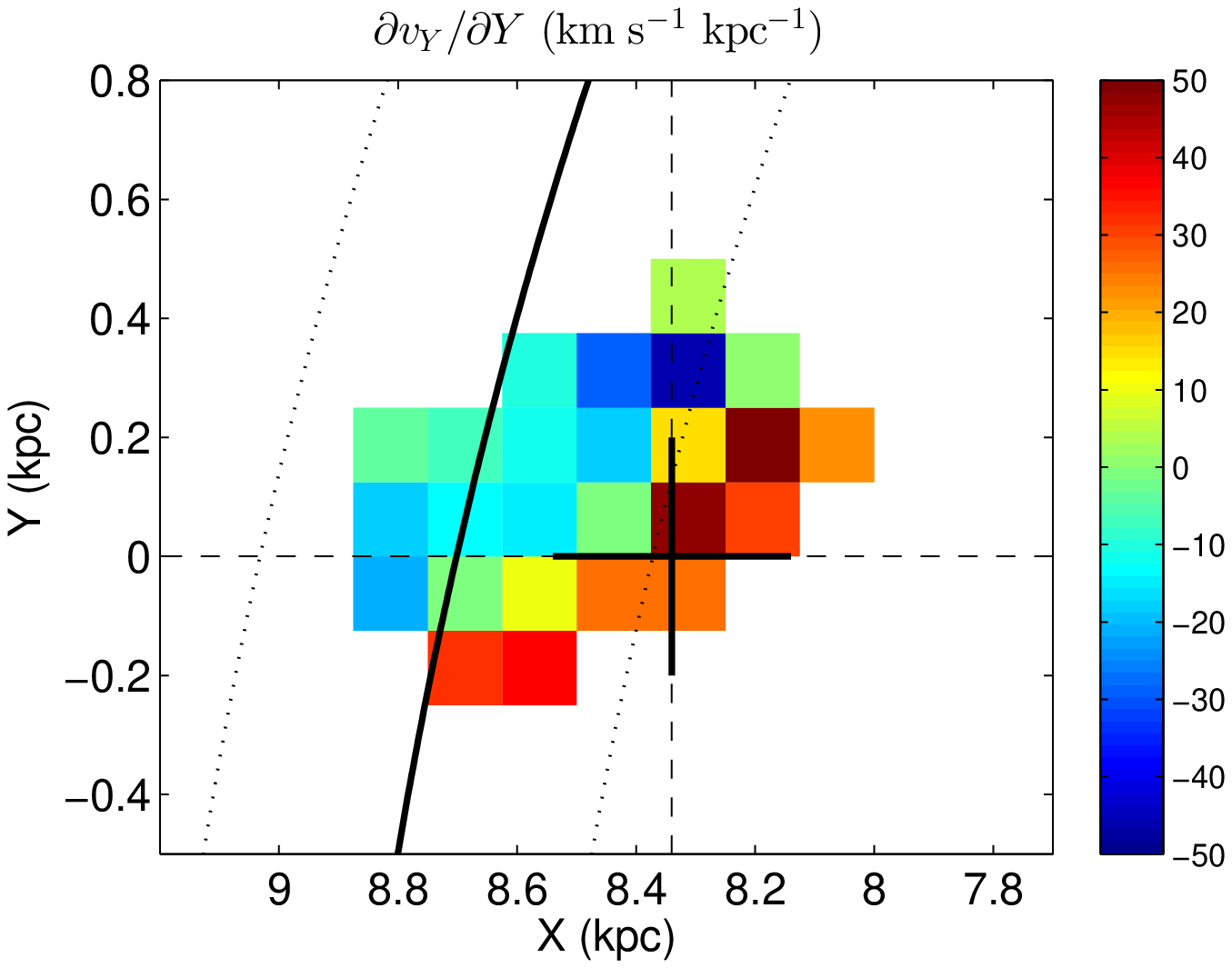}
\includegraphics[scale=0.5,trim={0cm 0 0 0},clip]{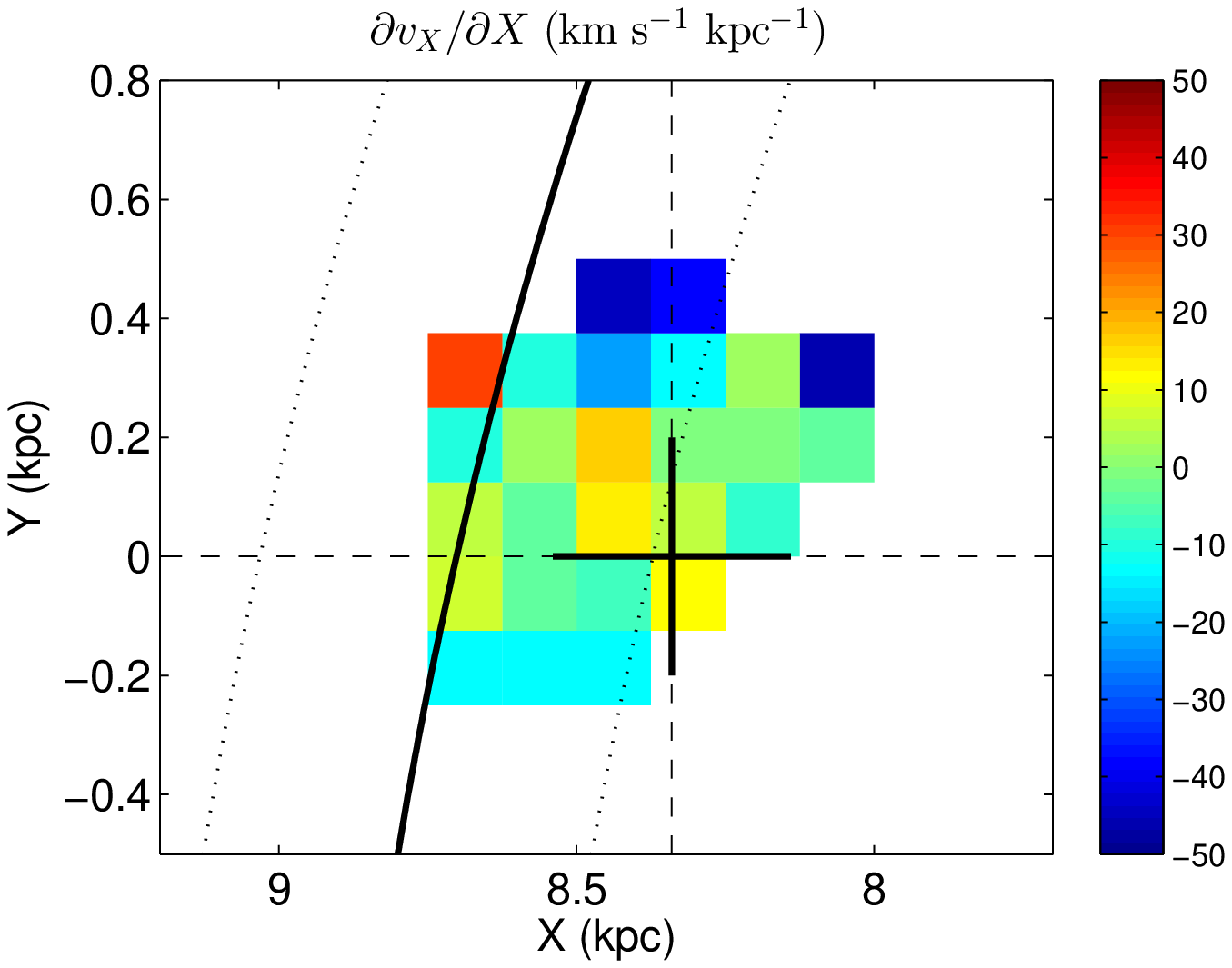}
\end{minipage}
\caption{In the top panel, the colors code the divergence of velocity in $Y$ direction, $\partial v_Y/\partial Y$. And in the bottom panel, the colors code the divergence of velocity in $X$ direction, $\partial v_X/\partial X$.}\label{fig:vectors}
\vspace{14pt}
\end{figure}

To better quantify the kinematical features, we display the two components of the divergence of the in-plane velocity in Figure~\ref{fig:vectors}. The top panel shows the map of $\partial v_Y/\partial Y$ and the bottom panel shows the map of $\partial v_X/\partial X$. First, the largest values of the divergence occurs in $\partial v_Y/\partial Y$, which varies from +30\,\kms\ kpc$^{-1}$ to -15\,\kms\ kpc$^{-1}$ from bottom-right to top-left in the $X$--$Y$ plane. Second, the direction of the variation of $\partial v_Y/\partial Y$ is roughly along the norm direction of the gas-identified local arm. Finally, $\partial v_X/\partial X$ is quite flat at around zero, implying that almost no gradient of velocity occurs along $X$.

Now consider the continuity equation of fluid mechanics, 
\begin{equation}\label{eq:continuity}
{d\rho\over{dt}}=-\rho({\partial v_X\over{\partial X}}+{\partial v_Y\over{dY}})-(v_X{\partial \rho\over{\partial X}}+v_Y{\partial \rho\over{\partial Y}}),
\end{equation}
if the stellar density does not rapidly change with time, then the left hand side of Eq. (\ref{eq:continuity}) essentially equals to zero. Because Figure.~\ref{fig:vectors} shows that the divergence of velocity is dominated by $\partial v_Y/\partial Y$ and the divergence along $Y$ ($X$) direction would only affect the variation of the density along $Y$ ($X$) direction, the variation of the stellar density $\rho$ should be also dominated by $\partial\rho/\partial Y$.

$\partial v_Y/\partial Y$ declines from +30\,\kms\ kpc$^{-1}$ to about $-15$\,\kms\ kpc$^{-1}$, passing through the zero point at the line from ($X$, $Y$)$\sim$(8.2, +0.3) to $\sim$(8.7, -0.1)\,kpc. Hence, $\partial\rho/\partial Y$ increases from negative values behind the line (w.r.t. the rotation direction) to positive values in front of the line according to Eq.~(\ref{eq:continuity}). Consequently, the stellar density $\rho$ must increase with $Y$ starting from the line. In other word, the young stars are associated with a density wave near the gas-identified local arm. Therefore, we suggest that the motion of the young stars probably reflects the local arm by kinematically associating themselves with a density wave. 

This scenario is not conflict with the mean age for the young tracers, which is only $\sim2$\,Gyr. Such a time scale is within the expected lifetime of the transient spiral structures~\citep{sellwood2014}.

\begin{figure*}[htb]
\centering
\includegraphics[scale=0.5,trim={0 0 0 1cm},clip]{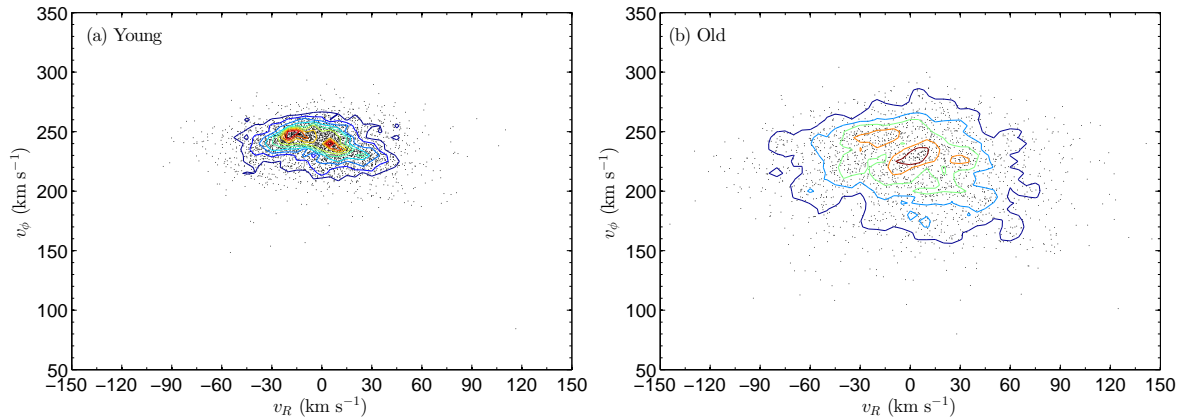}
\caption{Panels (a) and (b) show the distribution of the F-type young and K-type old SGB stars in \vPhi\ vs. \vR\ plane, respectively.}\label{fig:vrvphi}
\vspace{15pt}
\end{figure*}

Moreover, the distributions of the in-plane velocities displayed in Figure~\ref{fig:vrvphi} show that, for the young stars (panel a), the velocity distribution forms a tight arc shape. By contrast, the old stars (panel b) show an extended and roughly featureless distribution with a moderate bump at around \vR$\sim30$\,\kms\ and \vPhi$\sim180$\,\kms, which should be the Hercules stream~\citep{dehnen2000,xia2015}. The velocity distribution shown in panel (a) can be qualitatively compared with the simulations from~\citet{quillen2011}. We find that the orientation of the arc similar to our sample is located at the inter-arm regime (see the panels at the 3rd row and the 2nd, 5th, and 6th columns in their figure 8), not on the arm, which is consistent with the location of the majority of our samples.

It is noted that, other than the transient spiral mode~\citep{sellwood2014}, quite a few works have discussed about the alternative channels of the formation of the spiral structures and some theoretical works have given observational predictions in velocity distributions. Particularly about the local arm, \citet{xu2016} thought that it may be formed from the perturbation of the giant molecular clouds~\citep{donghia2013} according to their latest maser observations. And~\citet{li2016} found that the local arm, which can sustain over a short time, may not be due to the density wave, but be naturally induced in the scenario such that the disk contains two pairs of prominent spiral structures based on their SPH simulations. We also notice that, recently, \citet{hunt2016} claimed that the fast rotating stars found at the radii a few hundreds parsecs beyond the Sun may be driven by the Perseus arm other than the local arm. 

The different spiral theories may give different explanations to the peculiar motion of the young stellar populations unveiled in this work. Simply from the observed data, we cannot rule out other explanations. Further investigations, including more observations in larger volume and well defined N-body simulations, are anxiously required to clarify the nature of the peculiar velocity field for the young stars.

\section{Conclusions}\label{sect:conclusions}
In this letter, we study the in-plane peculiar motions using more than 3\,000 young stars combined the line-of-sight velocity from the LAMOST DR3 with the parallax and proper motion from the \emph{Gaia} DR1. The latter provides accurate proper motions with uncertainty of about 1\,mas yr$^{-1}$ up to 1\,kpc in distance, which is the best astrometric measurement so far. 

We select the F-type stars as the tracers for the young population and the old K-type SGB stars as the control sample. After removing the peculiar motions owe to the perturbations, the residual velocity field for the young stars show correlation with the gas-identified local spiral arm. The $\partial v_Y/\partial Y$ map for the young stars implies that the stellar density may increase with $Y$ starting from the line from ($X$, $Y$)$\sim$(8.2, +0.3) to (8.7, -0.1)\,kpc. This hints that the young stars are associated with a density wave located around the local spiral arm. We then suggest that the young stars are involved in the formation of the local arm by inducing the density wave and hence directly reflects the kinematical features of the arm. We also note that a few alternative mechanisms to explain such a peculiar velocity field are not simply ruled out. 

In future, it is worthy to directly compare the peculiar velocity field unveiled by the young stars in this work with various simulations so that the observational evidence can help to distinguish the origin of the spiral structures.

\acknowledgements
We appreciate the helpful comments from the anonymous referee. We thank Wyn Evens, Lia Athanassoula, and Ron Drimmel for their helpful discussions and comments. This work is supported by the Strategic Priority Research Program ``The Emergence of Cosmological Structures" of the Chinese Academy of Sciences, Grant No. XDB09000000 and the National Key Basic Research Program of China 2014CB845700. CL acknowledges the National Natural Science Foundation of China (NSFC) under grants 11373032 and 11333003. YGW acknowledges the NSFC grant 11390372.
This project was developed in part at the 2016 NYC Gaia Sprint, hosted by the Center for Computational Astrophysics at the Simons Foundation in New York City. 
Guoshoujing Telescope (the Large Sky Area Multi-Object Fiber Spectroscopic Telescope LAMOST) is a National Major Scientific Project built by the Chinese Academy of Sciences. Funding for the project has been provided by the National Development and Reform Commission. LAMOST is operated and managed by the National Astronomical Observatories, Chinese Academy of Sciences.
This work has made use of data from the European Space Agency (ESA) mission {\it Gaia} (\url{http://www.cosmos.esa.int/gaia}), processed by the {\it Gaia} Data Processing and Analysis Consortium (DPAC, \url{http://www.cosmos.esa.int/web/gaia/dpac/consortium}). Funding for the DPAC has been provided by national institutions, in particular the institutions participating in the {\it Gaia} Multilateral Agreement.

\end{document}